# Study of Pluto's Atmosphere Based on 2020 Stellar Occultation Light Curve Results

Atila Poro[1,2*], Farzaneh Ahangarani Farahani[1], Majid Bahraminasr[1], Maryam Hadizadeh[1], Fatemeh Najafi kodini[1], Maryam Rezaee[1], Mahsa Seifi Gargari[1]

[1]The International Occultation Timing Association Middle East section, Iran
[2]Astronomy Department of the Raderon Lab., Burnaby, BC, Canada

**Abstract**
On 6 Jun 2020, Pluto's stellar occultation was successfully observed at a ground-based observatory and Pluto's atmospheric parameters were investigated. We used an atmospheric model of Pluto (DO15), assuming a spherical and transparent pure $N_2$ atmosphere. Using ray-tracing code the stellar occultation light curve was satisfactorily fitted to this model. We found that Pluto's atmospheric pressure at the reference radius of 1215 km is $6.72 \pm 0.21$ μbar. Our estimated pressure shows a continuation of the increasing pressure studied in 2016 consistent with a seasonal volatile transport model. We concluded that the $N_2$ condensation processes in the Sputnik Planitia glacier are increasing due to the heating of the $N_2$ ice in this basin. This study's result was shown on the diagram of the annual evolution of atmospheric pressure.

**Keywords:** Occultations – Planets and satellites: atmospheres – Kuiper belt objects: individual: (Pluto)

## 1. INTRODUCTION

Pluto and its satellites are the best and most comprehensible things among Trans-Neptunian Objects (TNOs) due to their well-known complicated and active geological properties (Spencer et al. 2020). The Hubble Space Telescope observation revealed some features of this dwarf planet. Furthermore, NASA's New Horizons spacecraft made a close flyby of Pluto in 2015 in order to study Pluto and its moon Charon (Stern et al. 2015).

Since the confirmation of the existence of Pluto's atmosphere based on a 1988 stellar occultation (Hubbard et al. 1988), the study of Pluto's atmospheric parameters (pressure, composition, temperature, etc.) has carried on using data from both ground-based and space-based observation.

To calculate Pluto's atmospheric pressure, Elliot & Young (1992) developed a model; the model was based on a thermal gradient indicated in a light curve as a scale height. Elliot's model is a method for precisely coordinating the data obtained from an occultation light curve. The structure of Pluto's atmosphere was determined in the areas examined by the occultation. Additionally, the conceivable physical conditions of the atmosphere on the supposition were investigated (Elliot et al. 1989). Later in 2015, a simple atmospheric model was defined based on using both direct ray-tracing and inversion approaches by assuming a spherically symmetric, clear, and pure $N_2$ atmosphere (Dias-Oliveira et al. 2015). This model was suitably fitted with light curves from 2012 and 2013 stellar occultations between the heights of 1190 km and 1450 km from Pluto's center (Dias-Oliveira et al. 2015).

Research conducted during recent decades indicated Pluto's atmospheric pressure changes due to seasonal cycles of Pluto's surface volatiles which were calculated using atmospheric models (Meza et al. 2019). In 1998, atmospheric pressure at a radius of 1215 km (the distance to Pluto's center) was estimated at $2.33 \pm 0.24$ μbar (Yelle & Elliot 1997). Atmospheric pressure increased to $6.05 \pm 0.32$ μbar in 2008 (Sicardy et al. 2011). A drop in Pluto's atmospheric pressure was reported from 2008 to 2010, estimated at $5.64 \pm 0.22$ μbar (Young et al. 2010). The Stratospheric Observatory for Infrared Astronomy (SOFIA) aircraft, simultaneously with ground-based observatories, observed the 2015 stellar occultation by Pluto at an altitude of ~40,000 ft. (Bosh et al. 2015). Analyzing the data from SOFIA in optical and near-infrared wavelengths, along with ground-based observations, showed that the atmospheric pressure at half-light altitude was stable from 2011 to 2015 (Bosh et al. 2015). In 2015, Pluto's atmospheric pressure was observed at its maximum value of about $6.92 \pm 0.07$

---

µbar (Sicardy et al. 2016). According to a stellar occultation observation in 2019, Pluto's atmospheric pressure was estimated to be $5.20^{+0.28}_{-0.19}$ µbar which showed a decrease of approximately 21% between 2016 and 2019 at the 2.4σ level (Arimatsu et al. 2020).

In this study we observed a stellar occultation by Pluto on 06 Jun 2020. This observational data was used to obtain the atmospheric pressure of Pluto. Furthermore, the result and discussion about the obtained parameters are presented.

## 2. OBSERVATION AND DATA REDUCTION

The occultation of the star Gaia DR2 ID: 6864932072159710592, with an astrometric position of RA: $19^h$ $45^m$ $33.8957^s$ Dec: $-22°$ $10^m$ $19.0471^s$ (J2000), and $G$ magnitude of 12.9787[1] occurred on 06 Jun 2020 (UT). According to the prediction by the Lucky Star project[2] the observation was made with equipment of a private Asal observatory (Lat. 35° 56′ N, Lon. 50° 58′ E, Alt. 1947 m) in the north of Karaj City, Iran. To improve observing circumstances, a place was identified at a height not far from the observatory, and the observatory's equipment was transferred there. Therefore, the observation done in the geographical coordinates 35° 51′ N and 51° 01′ E and 2220 meters above the mean sea level and the horizon was free of any obstructions. This event was recorded with a 24-inch Schmidt-Cassegrain as a main telescope and a 14-inch Schmidt-Cassegrain telescope installed on the main telescope; both telescopes were on a Paramount equatorial mount. This observation was accomplished using a clear filter. Photometric data were derived from two CCD SBIG 11000M cameras with a 2004×1336 pixel array and pixel length of 18µ on the both telescopes. During the occultation the exposure time to cycle ratio of 0.81 seconds (dead time= 0.93 s) and the average temperature of the CCD was -25°C. The CCDs took images at exactly the same time under the same conditions. All the simultaneous images were merged in pairs to get an integrated image. The raw images were aligned by AstroImageJ software (Collins et al. 2017). DeepSkyStacker software (Ashley 2015) was used to merge each pair of simultaneous images. This image processing was necessary based on the low altitude of the event, the exposure time for each telescope's image (4s), and a large amount of light reduction of ~1.9 magnitude. A total of 234 images were taken of which 24 usable images were at the time of the occultation. It should be noted that some images have been deleted due to high noise or sudden atmospheric turbulence.

We reduced the images and corrected them. The data reduction was performed for bias, dark and flat field of each CCD image according to the standard method based on AstroImageJ software. Also, airmass was calculated based on the observation location that influenced and improved the light curve for the entire length of our observation. Accordingly, we computed the airmass based on the Astropy package in Python (Robitaille et al. 2013).

Figure 1 shows the Earth as seen from Pluto at the time of the occultation. The outer thin blue lines correspond to a predicted 1% drop in light from the star, the practical limit of detection, and the inner thin blue lines correspond to the zone where a half-light level was to be expected.

---

[1]http://vizier.u-strasbg.fr/
[2]https://lesia.obspm.fr/lucky-star/index.php

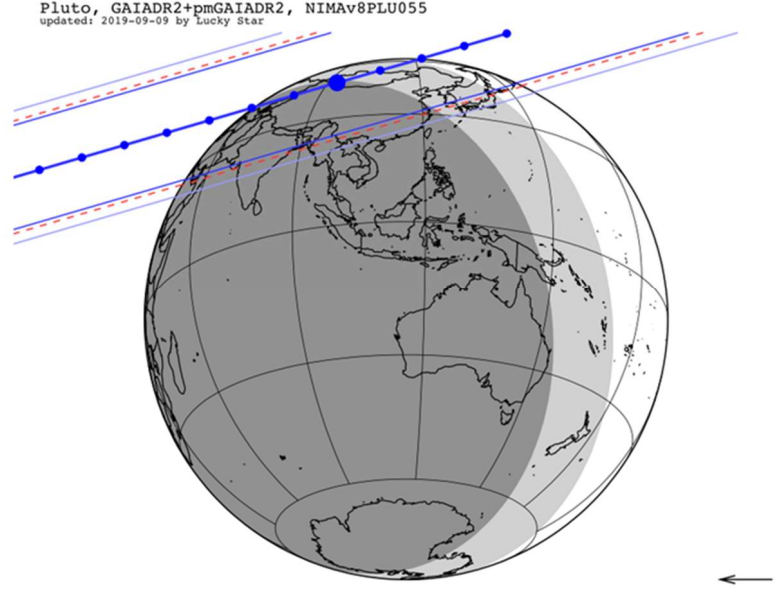

**Figure 1.** The path of Pluto's shadow on Earth on 6 June 2020. Small blue dots show track of central occultation geographically each minute before and after the predicted time of central occultation. The large dot indicates the location/time of central occultation in China. The arrow corresponds to the direction of the shadow motion. The precision of the red dashed line is the estimated 1-σ path deviation. 'When the Sun elevation is below -18 degrees (night), the areas are dark grey, and when the Sun elevation is between 0 to -18 degrees' areas are shown in light grey'. The prediction comes from the Lucky Star project.

## 3. METHOD OF ANALYSIS AND LIGHT CURVE FITTING

A best-fit of an atmospheric model was done to the occultation light curve data set. The extracted light curve is the total flux from the star and Pluto's system as a function of phase. The light curve after normalization is shown in Figure 2.

The atmospheric pressure of Pluto is a time-dependent parameter that has changed during the intervening years. Since the main component of Pluto's atmosphere is nitrogen ($N_2$), atmospheric pressure deviations have been caused by continuous $N_2$ condensation processes in the Sputnik Planitia glacier (Stern et al. 2015, Sicardy et al. 2016).

In order to obtain the pressure of Pluto's atmosphere, we fit the occultation light curve through a spherical and transparent atmospheric model of Pluto with a synthetic profile. This was created by using our ray-tracing code based on the pressure at the reference radius of $R = 1215$ km (Sicardy et al. 2016, Meza et al. 2019, Arimatsu et al. 2020) and relative Pluto+Charon flux as a reference value (Dias-Oliveira et al. 2015). The constant physical parameters that are used in this technique are given in Table 1. We used the same parameters as in Dias-Oliveira et al. (2015) and Meza et al. (2019).

**Table 1.** Physical constant and adopted parameters for the 6 June 2020 light curve fit and results.

| Parameter | Value |
|---|---|
| Pluto's mass ($GM_p$) | $8.69 \times 10^{11}$ m³/s² |
| Surface radius ($R_{surface}$) | 1187 km |
| Pluto's geocentric distance (D) | $4.9889 \times 10^9$ Km |
| Pluto pole position (J2000) | $\alpha_p$ = 08$^h$ 52$^m$ 12.94$^s$, $\delta_p$ = −06$^d$ 10' 04.8'' |
| $N_2$ molecular refractivity (K) | $1.091 \times 10^{-23}$ |
| Pressure at 1215 km ($P_{1215}$) | $6.72 \pm 0.21$ μbar |
| Pressure at the surface ($P_{surface}$) | $12.36 \pm 0.38$ μbar |

| Closest approach to Pluto's shadow center | 605.3 ± 5 km |
|---|---|

As in the DO15 atmospheric model described by Dias-Oliveira et al. (2015), all the equations depend only on the radius distance from Pluto's center. By assuming a spherically symmetric planet and a $N_2$ pure atmosphere in the DO15 model, we satisfactorily fitted our light curve. To demonstrate the quality of fit, the $\chi^2$ per degree of freedom has been obtained by a well-known relation,

$$\chi^2_{dof} = \frac{\chi^2}{N-M} = \frac{1}{N-M}\sum_{i=1}^{N}\frac{(\Phi_{i,obs}-\Phi_{i,syn})^2}{\sigma_i^2} \quad (1)$$

where for the $i^{th}$ data point, $\Phi_{i,obs}$ and $\Phi_{i,syn}$ are the observed and synthetic stellar fluxes respectively with σ error for each data point; $N$ is all the light curve data points and $M$ is the number of free parameters according to the model. The $\chi^2$ map as a function of the surface pressure, $P_{surface}$, and the distance of the closest approach of our observatory to Pluto's shadow center, $\rho$, is shown in Figure 3. According to the simultaneous fit to our occultation light curve, the $\chi^2$ value is 10.23 with 12 degrees of freedom (24 data points-12 free parameters). This achieved the satisfactory fit of $\chi^2_{dof} = 0.85$. Table 2 presents the $\chi^2_{dof}$ value for occultation events from 1988 to 2020.

The best-fitting parameters returned from our ray-tracing code are listed in Table 1. In particular, the reference radius pressure is about $6.72 \pm 0.21$ μbar (1σ level error bars), which corresponds to the surface pressure of $12.36 \pm 0.38$ μbar from fitting the DO15 model.

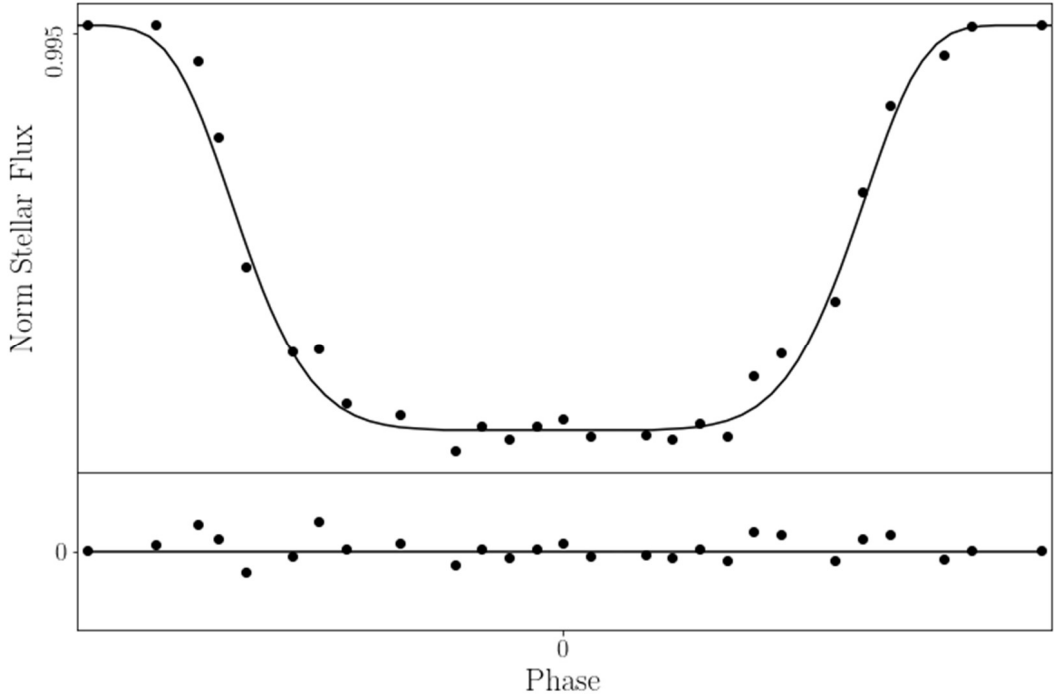

**Figure 2.** The normalized occultation light curve with the best-fit atmospheric model from the 6 June 2020 occultation. The solid curve shows a simultaneous fit to our light curve, using the DO15 atmospheric model. The residuals are also added at the bottom of the panel. The light curve also shows some of the data that was out of occultation. The zero phase time is based on the median time (19:04:01:35 UT) calculated in this study.

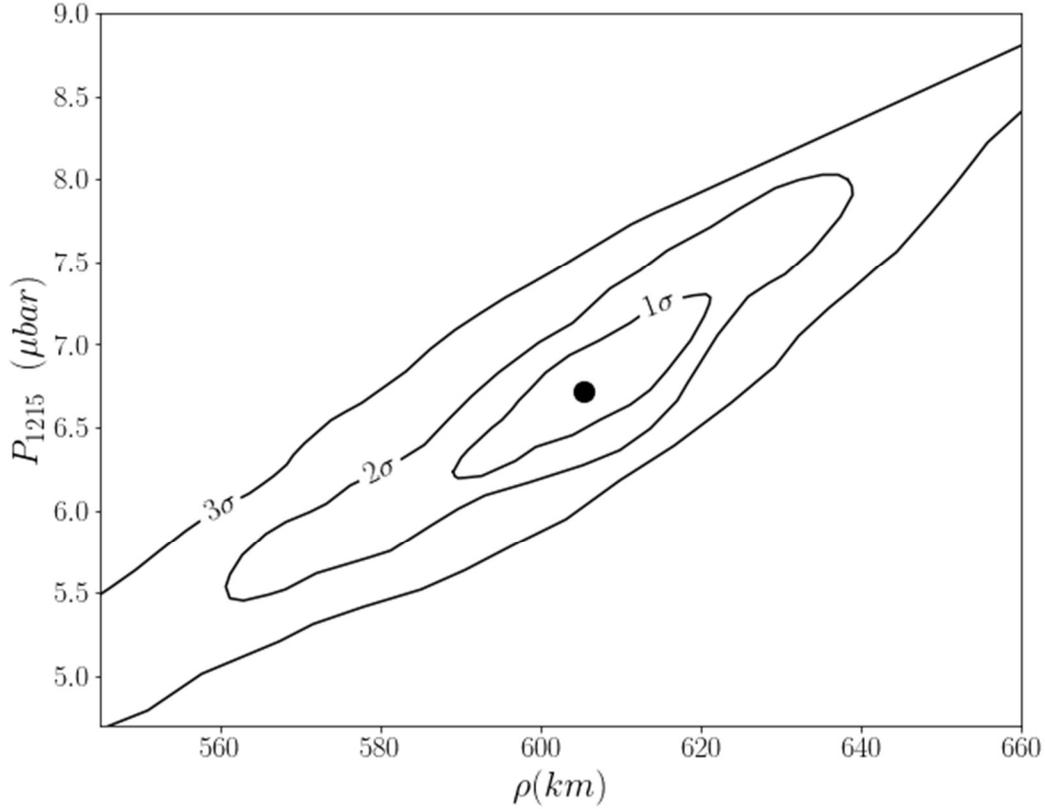

**Figure 3.** The $\chi^2$ map as a function of P$_{surface}$ and ρ derived from the simultaneous fit to our light curve. The minimum $\chi^2$ value of 10.23 with 12 degrees of freedom marks with a white circular point. The 1σ ($\chi^2_{min}$+ 1), 2σ ($\chi^2_{min}$+4), and 3σ ($\chi^2_{min}$+ 9) levels, are also shown. The 1σ curve provides the error bar.

In order to understand the atmospheric pressure evolution of Pluto, a numerical seasonal volatile transport model was developed by Bertrand & Forget (2016) at the Laboratoire de Météorologie Dynamique (LMD). According to this model and observed ground-based stellar occultations (Table 2), a peak is distinguished in the pressure values. Due to the maximum rate of the N$_2$ ice insolation in Sputnik Planitia's glacier, this increase in pressure is considered the main storage area for N$_2$ ice on Pluto. Results showed the seasonal Thermal Inertia (TI) between 500 and 1500 J $s^{-1/2}$ m$^{-2}$ K$^{-1}$, which is the principal parameter that can affect the N$_2$ cycle. The amount of N$_2$ in the atmosphere leads to different surface and atmospheric pressure. The LMD model defines that high TI leads to more formation of N$_2$ seasonal ice, low volatility, and an increase in atmospheric pressure. After this period, a decrease in pressure was predicted because of the decrease in the condensation of N$_2$ in Sputnik Planitia. These seasonal changes caused by the variation of obliquity and solar longitude of perihelion indicate the connection between Pluto's seasons and its distance from the Sun. The 2016 study simulated the nitrogen cycle over millions of years based on these changes. It investigated the rate of N$_2$ sublimation and condensation in different timescales according to the stability of N$_2$ ice in different longitudes (Bertrand et al. 2018).

In 2019 Meza and et al. constrained the LMD Pluto volatile transport model, derived based on a uniform seasonal TI of 800 J $s^{-1/2}$ m$^{-2}$ K$^{-1}$ and N$_2$ ice albedo range of $A_{N2} = 0.72$–$0.73$. As a result, the monotonic increase of pressure from 1988 to 2016 was consistent with this model (Meza et al. 2019).

**Table 2.** Recent Pluto's atmospheric pressure determined from ground-based stellar occultations.

| Observation | Surface pressure (μbar) | Pressure at the radius of 1215 km (μbar) | $\chi^2_{dof}$ | Reference |
|---|---|---|---|---|
| 1988 Jun 09 | 4.28 ± 0.44 | 2.33 ± 0.24 | NA | (Meza et al. 2019) |
| 2002 Aug 21 | 8.08 ± 0.18 | 4.42 ± 0.09 | 1.52 | (Meza et al. 2019) |
| 2007 Jun 14 | 10.29 ± 0.44 | 5.6 ± 0.24 | 1.56 | (Meza et al. 2019) |
| 2008 Jun 22 | 11.11 ± 0.59 | 6.05 ± 0.32 | 0.93 | (Meza et al. 2019) |
| 2008 Jun 24 | 10.52 ± 0.51 | 5.73 ± 0.21 | 1.15 | (Meza et al. 2019) |
| 2010 Feb 14 | 10.36 ± 0.4 | 5.64 ± 0.22 | 0.98 | (Meza et al. 2019) |
| 2010 Jun 04 | 11.24 ± 0.96 | 6.12 ± 0.52 | 1.02 | (Meza et al. 2019) |
| 2011 Jun 04 | 9.39 ± 0.70 | 5.11 ± 0.38 | 1.04 | (Meza et al. 2019) |
| 2012 Jul 18 | 11.05 ± 0.08 | 6.07 ± 0.04 | 0.61 | (Meza et al. 2019) |
| 2013 May 04 | 12.0 ± 0.09 | 6.53 ± 0.05 | 1.20 | (Meza et al. 2019) |
| 2015 Jun 29 | 12.71 ± 0.14 | 6.92 ± 0.07 | 0.84 | (Meza et al. 2019) |
| 2016 Jul 19 | 12.04 ± 0.41 | 6.61 ± 0.22 | 0.86 | (Meza et al. 2019) |
| 2019 Jul 17 | $9.56 \pm^{0.52}_{0.34}$ | $5.20 \pm^{0.28}_{0.19}$ | 0.84 | (Arimatsu et al. 2020) |
| 2020 Jun 06 | 12.36 ± 0.38 | 6.72 ± 0.21 | 0.85 | This study |

**4. RESULTS AND CONCLUSION**

Pluto's stellar occultation on 6 Jun 2020 was observed to obtain Pluto's atmospheric parameters. For the sake of finding and comparing the pressure at the specified height of the atmosphere, we satisfactorily fit our occultation light curve with the DO15 atmospheric model taking into account a spherically symmetric planet and $N_2$ pure atmosphere. Analysis of our single-chord photometry data demonstrates that the atmospheric pressure at the reference radius of 1215 km is $6.72 \pm 0.21$ µbar.

Pluto's atmospheric pressure values from 1988 to 2016 were retrieved by Meza et al. (2019). These were obtained based on the same light curve fitting model (DO15) except for the 1988 study (Table 2). Pluto's atmospheric pressure at the reference radius of 1215 km, P1215, is plotted as a function of time in Figure 4 to explain our occultation result compared to previous occultation light curves from the period 1988-2019.

As shown in Figure 4, the three-fold increase in pressure observed in 2015 is consistent with the expected pressure evolution from the LMD Pluto volatile transport model (Bertrand & Forget 2016). The presently derived pressure also is consistent with the continuous increase in pressure which has been observed since 1988. Based on this study's results, there is a small pressure decrease of 3% from the highest reported pressure value in 2015, and an increase since 1988 by a factor of three. The latest reported Pluto's stellar occultation by Arimatsu et al. (2020) in July 2019 showed a decrease in pressure rate of about 21% compared to the latest reported data from a stellar occultation in 2016. The Pluto volatile transport model predicts a decrease in surface pressure during the next few years due to the orbital decline of solar insolation and more condensation of $N_2$ ice in the colder part of Sputnik Planitia (Bertrand & Forget 2016). However, the obtained pressure drop by Arimatsu et al. (2020) was greater than the model predictions.

According to the stellar occultation observed and analyzed, this study's consistency with past studies indicates that the sublimation and condensation rate of $N_2$ ice in Sputnik Planitia's glacier is continuing to increase. According to the models an ascending increase was observed in atmospheric pressure since 1988 due to volatile ice dispensation (Bertrand & Forget 2016).

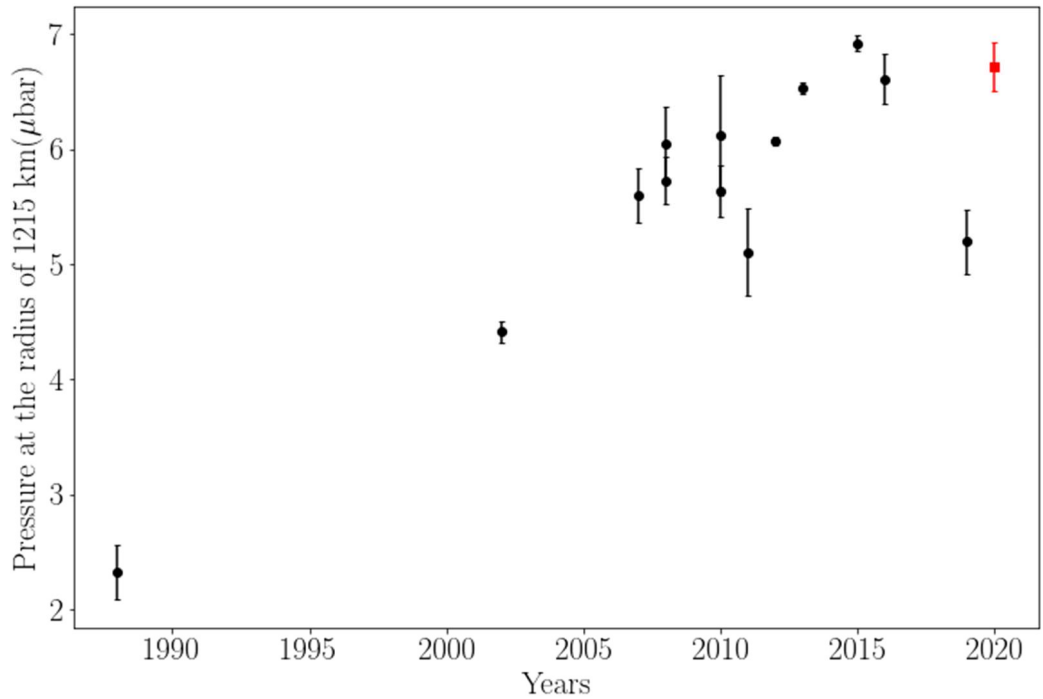

**Figure 4.** Pluto's atmospheric pressure at the reference radius of $R = 1215$ as a function of time. The red square dot is for this study. Black dots show the other atmospheric pressure obtained from 1988-2019 stellar occultation measurements with their error bars.

**ACKNOWLEDGMENTS**

The authors thank the ERC project n°669416 "Lucky Star" for making the prediction of this event publicly available. Furthermore, we are grateful to Paul D. Maley for making editorial corrections to the text and Fatemeh Hasheminasab for preparing the Latex version of the paper.